Persisting Roughness When Deposition Stops

Moshe Schwartz
School of Physics and Astronomy
Raymond and Beverly Sackler Faculty of Exact Sciences
Tel Aviv University
Ramat Aviv, Tel Aviv 69978 Israel

and

S.F. Edwards
Polymer and Colloid Group
Cavendish Laboratory
Cambridge CB3 OHE  UK

Abstract

  Useful theories for growth of surfaces under random deposition of material have been developed by several authors. The simplest theory is that introduced by Edwards and Wilkinson (EW), which is linear and soluble. Its non linear generalization by Kardar, Parisi and Zhang (KPZ), resulted in many subsequent studies. Yet both theories EW and KPZ contain an unphysical feature. When deposition of material is stopped both theories predict that  as time tends to infinity, the surface becomes flat. In fact, of course, the final surface is not flat, but simply has no gradients larger than the gradient related to the angle of repose. We modify the EW and KPZ to accommodate this feature and study the consequences for the simpler system which is a modification of the EW equation. In spite of the fact that the equation describing the evolution of the surface is not linear, we find that the steady state in the presence of noise is not very different in the long wave length limit from that of the linear EW. The situation is quite different from that of EW when deposition stops. Initially there is still some rearrangement of the surface but that stops as everywhere on the surface the gradient is less than that related to the angle of repose. The most interesting feature observed after deposition stops is the emergence of history-dependent steady state distributions.

In recent years there has been much activity in the study of the statistical properties of the evolution of surfaces when granular material is deposited. In the simplest model proposed by Edwards and Wilkinson (EW) [1], the spatial and temporal fluctuations of the surface are caused by random deposition followed by the diffusion of material to suppress gradients in the surface. The surface is described by a height function $h(\mathbf{r},t)$ above its mean and the EW equation for the evolution of the surface reads

$$\frac{\partial h}{\partial t} = \nu \nabla^2 h + \eta , \qquad (1)$$

where the noise term $\eta$, represents the local fluctuation in the rate of deposited material. The correlations of the noise are given by

$$\langle \eta(\mathbf{r},t)\eta(\mathbf{r'},t') \rangle = 2G(|\mathbf{r}-\mathbf{r'}|)\delta(t-t') . \qquad (2)$$

(The function $G$ must have, of course, a positive Fourier transform.) Usually, $G$ is taken to be a $\delta$ function but to take into account that the deposited material consists of particles of a finite size, $G$ has to have a finite range corresponding to that size.

The EW approach clearly oversimplifies the description of the quite complex way that grains land and settle and extensions of the equation have appeared in the literature. Particular attention has been paid to the KPZ extension [2] where the effect of the existing surface on the deposition is modeled by

$$\frac{\partial h}{\partial t} = \nu \nabla^2 h + g\ (\nabla h)^2 + \eta \ . \qquad (3)$$

This equation contains new physics. For example, the non linear term appearing in equation (3) above could be clearly related to lateral growth and the formation of overhangs as described in detail in ref. [2], where the geometrical motivation of that term is given. KPZ is a well behaved equation with a resemblance to the Navier-Stokes equation for turbulence. Its steady state is exactly soluble in one dimension [3] but the equation is also tractable in the statistical sense in higher dimensions [4-10], where theory and simulations agree on the power laws of the surface roughness and time evolution.

When the EW and KPZ equations are considered as equations for the deposition of dry material, there is however a feature of both equations which is clearly unphysical. When deposition is taking place, arbitrary gradients appear which of course are washed away by diffusion. If the deposition is suddenly turned off, steep gradients, defined by the absolute value of the gradient being larger than the tangent of the angle of repose, $\gamma$, crumble away but gradients less than $\gamma$ survive. Indeed this is a familiar fact. After the sand storm, there are still dunes. Contrary to that, EW and KPZ always give a surface that becomes flat as time goes by, once the deposition ceases.

It should be noted that the original EW and KPZ forms may be adequate when the deposited material sticks to the surface. For such systems it may be argued that as deposition stops, the flattening mechanism is suppressed. Namely, the coupling strengths in front of the terms responsible for flattening vanish as deposition stops. Eventual flattening can happen only by dislodging the particles sticking to the surface. This can be affected by other particles impinging on the surface particles or their vicinity. Thus, as deposition stops the surface stops evolving. In fact this is the physical reason for the identification of the exponents describing metal surfaces measured after deposition has stopped, with EW or KPZ steady state exponents that correspond to the situation of continuous deposition [11-13].

In the case of dry non-sticking material the diffusion constant in the EW equation, for example, does not vanish even when the deposition stops, as it is mainly governed by gravity. The implication is thus that if EW or KPZ constituted an adequate description of granular deposition of dry non-sticking material, the surfaces generated by deposition would have to flatten after deposition stopped in obvious contradiction to our daily experience.

This paper describes how this situation can be corrected by modifying the EW and KPZ equations, in such a way that the surfaces, resulting after deposition ceases, have gradients bound from above but are not flat. We study the simplest version obtained by modification of the EW equation and find that in spite of its simplicity it still leads to non trivial and interesting analysis.

In constructing the model we take into account a number of considerations:
First, we expect the systems to support an angle of repose. This implies that the absolute value of the gradient has to be bound from above by some finite constant but only a long time after the deposition has stopped and the surface is static. Furthermore, in the absence of noise, any initial surface with gradients everywhere below the bound must be stable. We would also like the equation describing the dynamics of rearrangement to be local and involve the same sort of mechanisms leading to the EW or to the KPZ equations. A basic quantity in the physical description of surfaces growing under deposition is the current density of material rearranging itself on the surface. (This is a current density in the plain perpendicular to the direction of deposition.) In EW and KPZ the current density $j$ is proportional to $-\nabla h$. We expect however the local current density to vanish whenever $(\nabla h)^2 > \gamma^2$. We thus propose that the way to incorporate an angle of repose in one of the above equations is to make the replacement $\nabla h \to f((\nabla h)^2 - \gamma^2)\nabla h$, where $f(x)=0$ for $x \leq 0$ and tends to 1 for values of $x$ larger than some small positive $x_0$. The Edwards Wilkinson equation is replaced by

$$\frac{\partial h}{\partial t} = \nu \nabla \cdot [f((\nabla h)^2 - \gamma^2)\nabla h] + \eta \tag{4}$$

and the KPZ equation is replaced by

$$\frac{\partial h}{\partial t} = \nu \nabla \cdot [f((\nabla h)^2 - \gamma^2)\nabla h] + f^2((\nabla h)^2 - \gamma^2)(\nabla h)^2 + \eta. \tag{5}$$

It is clear that the above equations meet all our requirements. The equations are local and the current density $j$ vanishes whenever $(\nabla h)^2 \leq \gamma^2$. This implies that in the absence of deposition ($\eta = 0$), if everywhere $(\nabla h)^2 \leq \gamma^2$, the time derivative of the height is zero everywhere and the surface is static. Therefore, after deposition stops the surface will keep evolving not into a flat surface but rather into a surface in which the angles of the slopes are less than the angle of repose.

The equations above describe threshold dynamics that is introduced by the use of the soft $\vartheta$ function, $f$. Other kinds of threshold dynamics encountered in the study of general microscopic models of self organized criticality [14] and the evolution of river networks [15] are also naturally introduced by the use of the $\vartheta$ function.

We will concentrate in the following on the simplest model described by eq.(4) and start by discussing the steady state when the deposition is still on. Consider first the case where the noise correlations in eq.(2) are described by $G(\mathbf{r}) = \mathbf{D}\delta(\mathbf{r})$, The Langevin equation above can be replaced in a standard way by the corresponding Fokker-Planck equation for the distribution of the heights, $P\{h\}$,

$$\frac{\partial P}{\partial t} = \int d\mathbf{r} \frac{\delta}{\delta h(\mathbf{r})} \{D \frac{\delta}{\delta h(\mathbf{r})} - \nu \nabla \cdot [f((\nabla h)^2 - \gamma^2)\nabla h]\} P. \tag{6}$$

The steady state distribution is given by

$$P_S = N \exp\{-(\nu/2D)\int d\mathbf{r} F((\nabla h)^2 - \gamma^2), \tag{7}$$

where $N$ is the normalization constant and $f(x) = F'(x)$. The problem can be viewed as an equilibrium problem described by the "Hamiltonian"

$$H = (\nu/2) \int d\mathbf{r} F((\nabla h)^2 - \gamma^2). \tag{8}$$

(This is only a formal resemblance. The "Hamiltonian" does not have units of energy nor is temperature involved.)

There is no particular difficulty in replacing our general $f$ by its limiting form, the Heaviside $\vartheta$ function. $F(x)$ in eq (8) will be replaced by $\vartheta(x)x$. In the following we will thus use only the $\vartheta$ function which provides the simplest possible description. It is worth noting that the steady state in the specific case in which the $\vartheta$ function in the current density is combined with $\delta$ function correlations, is entirely equivalent to the EW steady state. The reason is that if an infinitely fine powder lands on a surface without any spatial correlations the size of the local gradient must be infinite everywhere, so that the argument of the $\vartheta$ function is always positive, resulting in a value of the $\vartheta$ function which is always one. (Indeed, it is easily shown that in a linear problem the probability to obtain $|\nabla h| \leq \gamma$ is proportional to $\gamma^2 / \langle (\nabla h)^2 \rangle$ and for an EW system $\langle (\nabla h)^2 \rangle$ tends to infinity with the high "momentum" cut-off, related to the finite size of the landing particles and is infinite for the $\delta$ function case.) We consider next the more physical case in which the noise correlations have a non vanishing range. Standard combination of symmetry considerations with scaling arguments [16] suggest that the model described above (eq.(4) Is in the universality class of the EW model. Namely, the exponents describing its small momemta ($q$) are identical to those of the EW model. We have calculated the structure factor $S_\mathbf{q} = \langle h_\mathbf{q} h_{-\mathbf{q}} \rangle$ in order to obtain its explicit dependence on the high "momentum" cut-off, $q_c$. The full derivation is beyond the scope of this paper. We present, however in the next equation the structure factor for a very large $q_c$. This will exhibit the dimensionless small parameter characterizing the model.

$$S_\mathbf{q} = \frac{D}{\nu}[1 + \frac{\gamma^4}{32\pi^2}(\frac{D}{\nu})^{-2} q_c^{-4}] q^{-2}. \tag{9}$$

The typical frequency, , $\omega_\mathbf{q}$ , giving the decay of a disturbance of wave vector $\mathbf{q}$, seems to give no problem either and to scale as in EW.

We turn now our attention to the regime in which the noise is turned off, which was the basic reason to introduce our modification of the EW model. First, we find it useful to introduce the effect of the finite size of the grains by considering a discrete version of eqs. ( 6-8) on a square lattice with lattice spacing a.

$$\frac{\partial}{\partial t}P = \sum_i (1/a^2) \frac{\partial}{\partial h_i}\{D\frac{\partial}{\partial h_i} + \frac{\partial}{\partial h_i}H\}P , \qquad (10)$$

where the "Hamiltonian" is given by

$$H = \frac{\nu}{2}\sum_i a^2 \vartheta((\tilde{\nabla}h_i)^2 - \gamma^2)((\tilde{\nabla}h_i)^2 - \gamma^2) \qquad (11)$$

and where $\tilde{\nabla}h$ is the discrete gradient, defined by its components along the axes

$$\tilde{\nabla}_j h_i = [h_{(i+\hat{j})} - h_{(i-\hat{j})}]/2a , \qquad (12)$$

$\hat{j}$, denoting a unit vector in the j direction. ( We use here only the $\vartheta$ function version.) Now consider a steady state characterized by the initial strength of the noise, $D_i$ and then reduce the strength of the noise to $D_f$. As long as $D_f > 0$, it does not matter whether the change is adiabatic or abrupt. At very long time the distribution tends to

$$P_f = N_f \exp[-H/D_f] , \qquad (13)$$

which is the only steady state distribution with finite $D_f$. The physical state without deposition is characterized by $D_f = 0$. The difference between finite and zero deposition is enormous. In the latter case the number of steady states distributions is infinite as deduced from the Fokker-Planck equation for the evolution of surface distribution (eq. (10)). Let's clarify this point. It is clear that in the absence of noise all height configurations $\{\bar{h}_i\}$, such that $|\tilde{\nabla}h| \leq \gamma$ everywhere, remain fixed in time. Each of these height configurations corresponds trivially to the height distribution $P_{\{\bar{h}\}} = \prod \delta(h_j - \bar{h}_i)$. Therefore contrary to the single steady state in the case of finite strength of the noise, In the case of zero noise there is a whole subspace of distributions spanned by the distributions $P_{\{\bar{h}\}}$, which is steady. All states in that subspace are fixed in time in the absence of noise. The time dependent distribution function in the absence of noise must thus tend as time tends to infinity to a state that belongs to that subspace. The fact that in the absence of noise the steady state is not unique, is the reason for the dependence of the long time distribution on initial conditions and on the way the noise is turned off.

Consider first the case in which the noise is turned off adiabatically. What is the distribution that can be expected at infinite time? The adiabatic turning off of the deposition is described by $D = D(t)$, a function of time that tends very slowly to zero. The adiabatic solution of eq. (10) for a time dependent strength of the noise is

$$P(t) = N(t)\exp[-H/D(t)] . \qquad (14)$$

This means that as time tends to infinity all the surfaces where somewhere $|\tilde{\nabla} h| \geq \gamma$ tend to have zero probability, while all the other surfaces tend to have equal probabilities.

Consider next the case where the initial distribution is a steady state distribution and the noise is turned off abruptly. Do we expect the final distribution to be that obtained in the adiabatic case? Furthermore, starting from two different steady state distributions, do we expect the final distributions to be identical? The answer to the first question is actually included in the answer to the second question. Now, the final state into which an initial distribution eventually evolves is the projection of that initial state on the space of zero noise steady states. It is not very probable though that two steady states characterized by different values of the strength of the noise, $D_{i1}$ and $D_{i2}$, have the same projection on the space of zero noise steady states. Yet it might perhaps happen as a result of some hidden symmetry. To rule that out it is worth while to try and understand the actual physical difference between final steady states with different initial steady states corresponding to different $D$'s. To answer that consider first what happens to a single surface when the deposition is turned off. The first point to observe is that it follows from the equation of motion

$$\frac{\partial}{\partial t} h_i = -\frac{\partial}{\partial h_i} H \ , \tag{15}$$

that $H$ is a monotonically non-increasing function of time. This is general and independent on $H$ and follows directly by considering the time derivative of $H$. A property that does depend on the model, is that on sites where initially $|\tilde{\nabla} h| \geq \gamma$, the value of $|\tilde{\nabla} h|$ tends to $\gamma$ as time tends to infinity. Due to spill over it may also happen that on some of the sites where initially $|\tilde{\nabla} h| < \gamma$ it will eventually tend to $\gamma$. In spite of the last effect, which confuses the issue, it is still clear that if we compare two initial surfaces $\{h_i\}$ and $\{\alpha h_i\}$ with $\alpha > 1$, we will find that the first surface will tend at infinity to a surface where the number of sites for which $|\tilde{\nabla} h| = \gamma$ is less than the number of corresponding sites for the second surface. The larger the value of the initial strength of the noise, the higher is the relative probability of finding initially higher slopes. Thus, if initially the value of $D$ is large enough it will allow that typically $|\tilde{\nabla} h| \gg \gamma$, so that eventually at infinite time $\langle (\tilde{\nabla} h)^2 \rangle = \mu(D) \gamma^2$, where $\mu(D) < 1$ but very close to it. If on the other hand the initial $D$ is very small, the infinite time average will be characterized by $\mu(D) \ll 1$. The same is true for a continuous system in which the finite size of the landing particles is represented by a high "momentum" cut-off rather than by discreteness of the system. The full picture is thus that the final local roughness contains information about the initial distribution. Larger local roughness in the final state indicates larger initial noise. Put in other words steeper sand dunes imply deposition by stronger and more turbulent winds.

**Acknowledgement** - We would like to thank Y. Kantor for reading the manuscript and for his most helpful comments.